\documentclass[prc,aps,preprint,showpacs,nofootinbib]{revtex4}
\usepackage{graphicx}

\begin{document}
\title{Shape Deformations in Atomic Nuclei} 

\author{ Ikuko Hamamoto$^{1,2,3}$ and Ben R. Mottelson$^{2,4}$ }

\affiliation{
$^{1}$ {\it Division of Mathematical Physics, Lund Institute of Technology 
at the University of Lund, Lund, Sweden} \\ 
$^{2}$ {\it The Niels Bohr Institute, Blegdamsvej 17, 
Copenhagen \O,
DK-2100, Denmark} \\ 
$^{3}$ {\it Riken Nishina Center, Wako, Saitama, 351-0198 Japan} \\ 
$^{4}$ {\it NORDITA, Blegdamsvej 17, 
Copenhagen \O,
DK-2100, Denmark} \\ 
}

\maketitle

\vspace{2cm}
The ground states of some nuclei are described by densities and mean fields 
that are spherical, while others are deformed. The
existence of non-spherical shape in nuclei represents a spontaneous symmetry
breaking.

\newpage

\section{INTRODUCTION}
The low energy spectra of many nuclei 
can be described by independent particle motion
in a mean field potential that has spherical symmetry, but the
spectra of many others 
are most simply described by a mean field that deviates significantly
from spherical symmetry.  Figure 1 presents the regions of neutron and proton
particle numbers ($N,Z$) 
where these deviations ({\it i.e.\/} deformations) have
been empirically observed.
The existence of deformation is made apparent by the related
physical phenomena: (i) Electric quadrupole moments\footnote{Electric 
quadrupole moment
is defined in eq. (\ref{eq:Q0}) and is a measure of the simplest 
deviation from spherical
symmetry in the nuclear density distribution.} 
and rates for electromagnetic quadrupole
transitions that are orders of magnitude 
larger than can be accounted for by the
estimates for quantum transitions of a single proton. (ii) The low
energy spectrum contains sequences of excited states with consecutive 
values of
the quantized angular momentum, $I$, with associated excitation energy varying
as $I(I+1)$ (as is known for the rotational kinetic energy of molecules).
(iii) In addition to the rotational excitations the spectra exhibit low-energy 
single-particle excitations that can be interpreted 
in terms of particles moving independently in a potential with spheroidal shape.  

The occurrence of deformation is a prerequisite to collective rotation of a
quantum system because otherwise one cannot speak of an orientation for the
system.  All molecules are deformed because they are built from a (small) number
of atoms which unavoidably break the rotational symmetry of the system. The
close connection of deformation and rotation is not only crucial in the
understanding of nuclear structure, but also at once allies this project with a
wide range of currently studied man-made finite systems as clusters of atoms,
quantum dots, and trapped cold atomic gases.

\section{Physics and geometry of broken rotational symmetry}
The nuclear many-body Hamiltonian involves kinetic energy that does not depend
on the orientation of the coordinate axes 
together with interactions that depend
on the relative coordinates of the particles and therefore is also invariant
under rotations of coordinate axes involving all of the particles at once.  
This rotational symmetry does not however exclude the possibility of finding 
that the lowest-energy mean field solution is obtained 
with a non-spherical mean potential and density distribution. 
The non-spherical solutions are interpreted as deformed intrinsic states.  
The occurrence of rotational spectra implies an approximate separation of the
collective rotation as a whole from the intrinsic motion in the body-fixed
system as described by a Hamiltonian  
\begin{equation}
H = H_{int} + H_{rot} 
\end{equation}
where the mean field of $H_{int}$ is non-spherical. 
Then, the total wave function $\Psi$ 
can be written, in the first approximation, as
\begin{equation}
\Psi = \phi_{int}(q) \, \Phi_{rot}(\Omega)
\label{eq:prowf}
\end{equation}
where the variables $q$ describe the motion of the individual neutrons and
protons with respect to an intrinsic (body fixed) coordinate system while 
$\Omega$ describes the orientation of the latter system with respect to 
the laboratory.  (See Figures 2 and 3.)
A general discussion of the form of $H_{rot}$ can be found in the Chapter 4 of
\cite{BM75} or in the section 3.3 of \cite{RS80}. 

The origin of nuclear deformations can be related to the shell structure of
single-particle levels ({\it i.e.\/} 
the bunching of single-particle energy levels) 
in a spherical potential. The occurrence of significant
gaps in the level spacing at the major shell closures provides the
rational for the special stability of these nuclei and at the same time
implies that the single-particle level density in the middle of the shell 
for a spherical mean field is appreciably
greater than the over-all average.  
All single-particle wave functions in the spherical potential (except for
$p_{1/2}$ and $s_{1/2}$) have a large intrinsic anisotropy and thus in the
presence of a deformed one-particle potential some single particle orbits will
move up in energy and others will move down as a function of deformation. 
This fanning out of the one-particle energies implies that the excessive
one-particle level
density in the middle of closed shells is relaxed.  
Thus, the filling of the lowest
orbits may provide an energy gain stabilizing the deformed configuration. 
The eccentricities obtained in this way for nuclear ground states are small
compared with unity, reflecting the fact that the
number of particles in open shells is small compared with the total 
($A = N + Z$). 
 
The existence of non-spherical shape in nuclei represents a spontaneous 
symmetry breaking and the rotational excitations are the Goldstone mode 
for this symmetry breaking.
The full rotational degrees of freedom\footnote{This implies $(2I+1)^2$
different rotational states for each value of $I$.} 
are obtained if the deformation
completely breaks the rotational symmetry (Figure 2).  However, if the
deformation is invariant with respect to a subgroup of rotations of the
coordinate frame, the collective rotational degree of freedom is 
correspondingly restricted.  For example, if the deformation is axially
symmetric, the rotation about the symmetry axis does not belong to collective
rotation, but is part of the intrinsic particle or vibrational 
degrees of freedom\footnote{There are now ($2I+1$) different rotational states
for each value of $I \geq \mid K \mid$ where $K$ is a constant intrinsic quantum
number characterizing the rotational band.} (Figure 3).

The symmetry of deformation restricts 
the quantum numbers of the rotational spectra.
Examples are ; (a) Space reflection invariance of deformation leads to
parity as a good quantum number in the rotational spectra, {\it i.e.\/} 
all members
in a given rotational band have the same space inversion parity, $\pi = \pm 1$. 
(b) Axially symmetric shape 
in quantum mechanics implies no collective rotation about the symmetry axis 
and that the intrinsic motion has a constant projection of angular momentum $K$ 
on the symmetry axis.  
Consequently, the projection, $K$, of the total angular momentum 
along the symmetry axis 
is a good quantum number which is common to all members of the rotational band. 
(c) When the deformation has $R$-symmetry ({\it i.e.\/} 
invariant with respect to 
a rotation of 180$^{\circ}$ about an axis perpendicular to the symmetry axis) 
in addition to axial symmetry, 
for example a deformation depending on the spherical harmonic 
$Y_{\lambda 0}$ with even $\lambda$, 
the $K$=0 states are eigenstates of the 
$R$-operator with eigenvalues $r = \pm 1$.  
The observed rotational bands comprise only the states 
\begin{equation}
I \: = \: \left\{ \begin{array}{ll} 0,~2,~4,~..... & \qquad 
\mbox{for} \quad K=0, ~~ r=+ \\
\\
1, ~3, ~5, ~..... & \qquad \mbox{for} \quad K=0, ~~ r=- 
\end{array} 
\right.
\label{eq:IK0}
\end{equation}
\begin{equation}
I \: = \: K, ~K+1, ~K+2, ~..... \qquad \mbox{for} \quad K \neq 0
\label{eq:IK}
\end{equation}
implying uniquely $R$-symmetry for these deformed nuclei. 
In the case of $K \neq 0$ the total wave functions take the form 
\begin{equation}
\Psi_{KIM} = \left( \frac{2I+1}{16 \pi^2} \right)^{1/2} \left( \phi_{K} (q)
D^I_{MK} (\Omega) + (-1)^{I+K} \phi_{\bar{K}} (q) D^I_{M,-K} (\Omega) \right)
\label{eq:intrwf}
\end{equation}
in order to fulfill the condition that the rotation by 180$^{\circ}$ 
about an axis
perpendicular to the symmetry axis is part of the intrinsic degrees of freedom
and should not be included in the rotational degrees of freedom \cite{BM75}. 
In Eq. (\ref{eq:intrwf})
$M$ denotes the component of ${\bf I}$ along the z-axis of a coordinate
system fixed in the laboratory, while 
the intrinsic state $\phi_{\bar{K}}(q)$ is the time-reversed state of
$\phi_{K} (q)$ or $\phi_{\bar{K}} (q) \equiv R_{i}^{-1} \phi_{K} (q)$
where $R_{i}$ expresses the operator $R$ acting on 
the intrinsic variables. 
The rotation matrices $D^{I}_{MK}$ depend on the orientation of the intrinsic
coordinate system with respect to the laboratory and are given in the standard
textbooks \cite{Rose57}.

The symmetrization (\ref{eq:intrwf}) of the $K \neq 0$ states leads to 
additional
terms in the rotational energy and intensity rules for these states.  For
example, for $K$=1/2 the rotational energy to leading-order in the angular
momentum acquires an additional term 
\begin{equation}
E_{rot} \: = \: A \left( I(I+1) + a \, (-1)^{I+1/2} \, (I+\frac{1}{2}) \, 
\delta(K, \frac{1}{2}) \right)
\label{eq:roten}
\end{equation}
where the parameter, $a = -<K=1/2 | j_+ | R_i^{-1}(K=1/2) >$, 
is referred to as the decoupling parameter 
while
$\bf j$ is the angular momentum of the single particle in the $\Omega$=1/2 
state if the $K$=1/2 can be associated with a single particle moving in the
potential generated by the rotating core.
Writing $A=\hbar^{2}/2\Im$, $\Im$ is the effective moment of inertia 
of the band.

In contrast to the symmetry, 
the size of deformation is related to the magnitudes of 
rotational energy (or moment of inertia) and E2 transition probability 
(or intrinsic quadrupole moment).

With only a single exception 
the observed rotational states (see for example Figures 4, 6 and 8) exhibit 
only the subset of the states identified 
in (\ref{eq:intrwf}), which uniquely identifies the existence of both axial- 
and $R$-symmetry
for these states, and   
in the rest of this article we shall confine
the discussion to this case; for a more general discussion see Ref. \cite{BM75}.
For the only known example, so far, of a nucleus clearly exhibiting triaxial
shape, see wobbling mode in Section VII.

\section{Moment of inertia for nuclear rotation}
The theory for the moment of inertia is based on the analysis of the extra
kinetic energy that is generated when the self-consistent deformed mean-field
potential is adiabatically 
set into uniform rotation with frequency $\vec{\omega}_{rot}$ 
about an axis perpendicular to the intrinsic symmetry axis.  
The Coriolis force, 
which is 
associated with this rotation and couples intrinsic motion with rotation, 
generates the moment of inertia as the second
order perturbation effect \cite{DRI54} 
of the single-particle angular-momentum operator, $j_x(k)$, 
acting on the
non-rotating ground state, $\mid 0 \rangle$.
\begin{equation}
\Im_x = 2 \hbar^2 \sum_{i \neq 0} \sum_{k=1}^{A} \frac{\langle i | j_x(k) | 0
\rangle^2}{E_i - E_0} 
\label{eq:crmt}
\end{equation}
where $i$ labels the many-particle states of the $A$ nucleons in the deformed
potential, with excitation energies $E_i$. 
For non-interacting nucleons moving in the deformed mean field, the
semi-classical evaluation of (\ref{eq:crmt}) leads, most amazingly, 
to a moment of
inertia equal to the value obtained for rigid rotation of a body with the given
mass density distribution\footnote
{The result, $\Im_x \rightarrow \Im_{x,rig}$, 
is based on a classical argument first found in
Niels Bohr's doctoral dissertation \cite{NB11} and usually referred to in the
condensed matter literature as the Bohr-van Leeuwen theorem. Since the
first-order effects of the rotation are equivalent to the effect of a magnetic
field, the absence of an induced flow in the rotating coordinate system
corresponds to the absence of diamagnetism in a classical electron gas.  In the
case of an anisotropic harmonic oscillator potential the rigid-body value of
non-interacting nucleons is just obtained for the equilibrium deformation
independent of configurations.  (The closed-shell configurations are a singular
exception to this result, since they have spherical equilibrium shape and a
vanishing moment of inertia.)  For other potentials one obtains fluctuations
of the moment of inertia about the rigid-body value depending on configurations 
\cite{BM55}.}, 
which is to leading order in the deformation 
\begin{equation}
\Im_{x,rig} \: = \: A \, M \, \langle y^2 + z^2 \rangle \:   
= \: \frac{2}{5} \, A \, M \, R^{2} \, \left( 1 + \frac{1}{3} \delta 
\right)
\label{eq:rigmt}
\end{equation}
assuming axial symmetry about the intrinsic z-axis, where $R$ is the mean
radius, $A$ denotes the mass number, $M$ the nucleon mass, while 
$\delta$ is the deformation parameter determined from the electric
quadrupole moment in (\ref{eq:Q0}).

Moments of inertia observed in the ground state bands of well-deformed even-even 
nuclei are systematically a factor of 2-3 smaller than 
the values for rigid rotation of the deformed body ($\Im_{rig}$), 
revealing a significant violation of the assumed independent particle motion of
the single particles.  The correlations revealed by the effect have been
understood in terms of a pair binding of the same kind as that involved in the 
Bardeen, Cooper, and Schrieffer (BCS) theory of 
superconductivity\footnote{The concepts developed for the treatment of
superconductivity in terms of correlations in the electronic motion 
\cite{BCS57}
provided a basis for analyzing the pair correlation in nuclei \cite{BM58}. 
See also Ref. \cite{STB59}}. 
The pair gap, $\Delta$, in the BCS theory is identified with the pair binding
observed in the systematically larger binding energies of nuclei with even
number of neutrons or protons as compared with the neighboring nuclei having an
odd number of the same (see the scholarpedia article on pair correlations in
nuclei).  The presence of the pair gap increases the energy denominators and
reduces the matrix elements in the numerators in the cranking calculation of the
moments of inertia (\ref{eq:crmt}) for the ground states of even-even nuclei and
reproduces rather well the main trends in the empirical moments for the
even-even nuclei, see Ref. \cite{NP61}.

Although the BCS correlations appear to provide the beginning of an
understanding of the nuclear moments of inertia, the particle-rotational
coupling continues to challenge our understanding since in the odd-A nuclei 
the same Coriolis     perturbation that generates the moments of inertia couples
near-lying rotational bands with $\Delta \Omega = \pm 1$, where $\Omega$ 
represents 
the component of the total angular momentum along the symmetry axis, 
and produces observable
effects in the rotational energies and E2 transition probabilities.  The
observed perturbations are systematically smaller by an appreciable factor than
for uncorrelated single particle motion, but in this case the BCS pair
correlations provide only a minor reduction (10-20 percent) leaving the observed
perturbation unexplained.
This "Coriolis reduction"
problem is still an open issue.\footnote{See F. Stephens (1960) quoted in Ref.
\cite{BM75}. Detailed discussions can be found 
in p.250-253, p.279, p.314 and p.318 of Ref. \cite{BM75}.}

\section{Deformation and rotational band in the $SU_3$ model}
The variables that describe the collective rotational motion are in general 
complicated functions of both the particle positions and momenta. 
Taking the particle motion within one-major shell of a harmonic oscillator
potential, the relationship between collective rotation 
and individual-particle motion
is clarified in the $SU_3$ model \cite{JPE58}.  
In the historical development the $SU_3$ model
was a wonderful and liberating contribution towards understanding the manner in
which interacting particles within a single major shell could produce
deformation and associated rotational band structure 
in the excitation spectrum.

For a system with $n$ nucleons the eight $SU_3$ operators consist of three
rotation operators
\begin{equation}
L_q = \sum_{k=1}^{n} \left( {\bf r}(k) \times {\bf p}(k) \right)_q
\end{equation}
and five quadrupole operators
\begin{equation}
Q_q^{(2)} = \sqrt{\frac{4 \pi}{5}} \sum_{k=1}^{n} \left( r(k)^2 \, 
Y_q^{(2)}(\theta_{r}(k), \phi_{r}(k) ) + b^{4} \, 
p(k)^{2} \, Y_q^{(2)}(\theta_{p}(k),
\phi_{p}(k) ) \right) \frac{1}{b^2}
\end{equation} 
where $b$ is the radial length parameter of the harmonic-oscillator 
wave functions 
while the arguments of the spherical harmonics are the polar angles of the
vectors ${\bf r}$ and ${\bf p}$, respectively. 
The oscillator Hamiltonian, $H_0 = r^2 + b^4 p^2$, is invariant with respect to
the above eight operators. The presence of the momentum-dependent term in 
$Q_q^{(2)}$ ensures that there is no mixing of different oscillator
configurations. 
Single-particle wave functions belonging to given irreducible representations of
the $SU_3$ group are eigenfunctions of $H_0$ as well as being eigenfunctions of
$Q_0$.  Thus, they may be regarded as approximations to the wave functions of a
deformed oscillator potential obtained by neglecting the mixing of different
oscillator shells $N$.  
Noting that the Casimir operator which is 
quadratic in the group operators is written as 
$C = \frac{1}{4} ({\bf Q}^{(2)} \cdot {\bf Q}^{(2)})_0 + \frac{3}{4} ({\bf L} 
\cdot {\bf L})_0$, the eigenvalues of the quadrupole-quadrupole force within the
$SU_3$ multiplet form a set of rotational bands with energies proportional to
$L(L+1)$.  

It is possible to construct the wave functions, 
which may be interpreted as intrinsic wave functions, for a given irreducible
representation which in general includes several intrinsic wave functions. 
Then, the
set of wave functions which are obtained from a given intrinsic state by
projecting all possible values of angular momentum forms an $SU_3$ multiplet. 
The multiplet with the greatest deformation lies lowest in energy with an
attractive quadrupole force.  
It is important to notice the presence of a particle-hole symmetry in the $SU_3$
model, which leads to the number of prolate systems equal to that of oblate
ones for the nuclear ground states.  
One expects prolate intrinsic shapes in the first half-shell and oblate
intrinsic shapes in the second half.  
If the electric quadrupole operator is proportional to the
group operator ${\bf Q}^{(2)}$, it has non-vanishing matrix elements only
between states in the same $SU_3$ irreducible representation.  Such transitions 
exhaust all the strength of the E2 transitions.  

Thus, the $SU_3$ classification provides an exact microscopic model in which a
quadrupole-quadrupole force leads to a ground state with large quadrupole
moment.  Moreover, not only the energy spectrum but also quadrupole 
transition probabilities have rotational characteristics in the limit of many
particles and not near the top of the band, where there is a band
termination accompanied by a smooth decrease in the E2 transition matrix-element
reaching $B(E2;I_{max} \rightarrow I_{max}+2)$ = 0 where $I_{max}$ is the
largest angular momentum in the band.
Since the maximum angular momentum of the single-particle and the number of
particles in the last filled shell are the order of $A^{1/3}$ and $A^{2/3}$,
respectively,
the rotational bands terminate at an angular momentum 
of order $A$ reflecting the finite dimensionality of the shell model space
involved.   
Having the results of the $SU_3$ model in mind, in the following section we
build the concepts of deformation and rotation directly from broken rotational
symmetry and a resulting intrinsic state.

\section{Single-particle states in deformed nuclei}
In the regions of particle number where the even-even nuclei exhibit rotational
band structure (Figure 1), the low energy spectra of the neighboring odd-A
nuclei can be interpreted in terms of rotational sequences based on the
different single particle or single hole states in the Fermi sea of particle
orbits in an appropriately chosen spheroidal potential.  Expressing a projection
of the angular momentum of the last odd particle by $\Omega$, 
observed values of
$K \pi$ for the lowest bands are equal to respective values of 
$\Omega \pi$, and the wave function of the odd particle is conveniently
used for the intrinsic wave function $\phi_{K}(q)$ in (\ref{eq:intrwf}); 
Examples are provided by observed spectra in Figures 6 and 8 which are
interpreted on the basis of the theoretical spectra in Figures 7 and 9.

The rotational coupling scheme between intrinsic and rotational angular momenta
is confirmed not only by the sequence of spin values and 
regularities in the energy spectra, but also by the intensity relations that
govern the transitions between states within a given band as well as those
between two rotational bands \cite{BM75}.
The leading order intensity rules are of a purely geometrical character
depending only on the rotational quantum numbers and the multipolarity of the
transitions.  

The starting point for the description of the intrinsic degrees of freedom 
is the analysis of single-particle motion in non-spherical potentials with the
symmetry and shape exhibited by the observed rotational spectra. 
The theoretical energy spectrum of one-particle eigenvalues 
plotted as a function of the axially-symmetric quadrupole deformation parameter 
is referred to as the 
Nilsson diagram \cite{SGN55}. 
The presence of a significant pairing correlation in the nuclear system (as
observed in the odd-even mass difference) implies that the low energy odd-A 
spectra are especially simple and for deformed nuclei can be completely
interpreted in terms of a single quasiparticle occupation of 
the Nilsson diagram orbits 
lying in the neighborhood of the Fermi energy. 

There are several definitions which have been conventionally used for the
parameter of axially-symmetric quadrupole ($Y_{20}$)
deformation.  Assuming a collective deformation of the nucleus as a whole, 
$\delta$ is defined using the intrinsic electric quadrupole moment 
\begin{equation}
Q_0 = \sum^{Z}_{k=1} \, \langle 2 \, z_k^2 - x_k^2 -y_k^2 \rangle 
= \frac{4}{3} \, \langle \sum^{Z}_{k=1} r_k^2 \rangle \, \delta 
\label{eq:Q0}
\end{equation}
while $\beta$ is defined in terms of the expansion of the radius
parameter 
\begin{equation}
R(\theta, \varphi) = R_0 \, (1 + \beta \, Y_{20}^*(\theta) + .....) 
\end{equation}
where $R_0$ is obtained by assuming a constant density for the undistorted 
spherical shape.  
To leading order, one obtains $\beta = \delta \sqrt{\frac{16 \pi}{45}} \approx 
1.06 \, \delta$.

The single-particle spectrum for an axially-symmetric
harmonic-oscillator potential  
\begin{equation}
V_{def.h.o.} = \frac{M}{2} \left( \omega_z^2 z^2 + \omega_{\perp}^2 (x^2 + y^2)
\right)
\label{eq:dhopot}
\end{equation}
is plotted in Figure 5.  
Note that in the potential (\ref{eq:dhopot}) 
there is neither surface nor spin-orbit potential, though both of these 
are important
in actual nuclei.  The lowest (highest) orbit for prolate
(oblate) shape is doubly degenerate ($n_{\perp} = 0$) and has an
axially-symmetric density distribution, which is built by taking a linear
combination of the one-particle orbits degenerate for spherical shape, while 
the density distributions of 
other one-particle orbits are in general not axially-symmetric.  
In the absence of two-body interactions the 
configurations with an unfilled shell in the harmonic oscillator potential have
always a deformed equilibrium shape. 
For example, the two-particles (two-holes) configuration in a given
$N \neq$ 0 shell of the harmonic oscillator potential has an equilibrium
deformation of prolate (oblate) shape.

It is seen that the strongest bunching of levels in Figure 5  
occurs for spherical shape but a similar order of 
level bunching occurs for large 
deformations with simple rational ratios of $\omega_{\perp} : \omega_z$. 
For large deformations such as 2 : 1 ($\delta_{osc} = 0.6$) or 1 : 2 
($\delta_{osc} = -0.75$) a shell structure which is essentially different
from that for spherical shape appears.  These deformations have not been
observed in the ground state of heavy nuclei, 
but observed fission isomers in actinide
nuclei \cite{AM73} and "superdeformed" yrast bands \cite{JK91} 
at high spins of some medium-heavy nuclei 
are understood in terms of 
the prominent shell structure and the associated magic numbers that occur for
$\omega_{\perp} : \omega_z = 2:1$ (superdeformation) 
in the harmonic oscillator potential. 

An example of observed well-studied low-energy spectra of 
deformed odd-A nuclei is shown in Figure 6.
The observed spectra can be understood in terms of the "aligned" coupling 
scheme, in which the orbital motion of nucleons is aligned with respect to the
orientation of the deformed field as illustrated in Figure 3.   
In this example the ground state
configuration ($K^{\pi} = 0^+$) of $^{24}_{12}$Mg$_{12}$ 
is characterized by the pairwise
filling of the time-reversed orbits aligned with respect to the symmetry axis
while the 13th neutron of $^{25}_{12}$Mg$_{13}$ 
moves in the spheroidal potential provided by
the even-even core of $^{24}$Mg.   Both energies and
$I^{\pi}$ quantum numbers of the observed states are successfully classified in
terms of rotational bands characterized by the quantum numbers $K^{\pi}$, which
correspond to the values of $\Omega^{\pi}$ of the odd neutron for a prolate
deformation with $\beta \approx 0.4$.  An example of the corresponding 
Nilsson diagram based on a
spheroidal Woods-Saxon plus spin-orbit potential is shown in Figure 7. 

In the classification of observed levels shown in Figure 6 
we have used the fact
that the levels (with $\Delta I \leq 2$) belonging to the same rotational band 
must be connected by strongly enhanced (collective) E2 transitions.   
Using the wave function in (\ref{eq:intrwf}) the reduced E2 transition
probabilities within a band are written as 
\begin{equation}
B(E2; KI_1 \rightarrow KI_2) = \frac{5}{16 \pi} \, e^2 \, 
Q_0^2 \, \langle I_1K20|I_2K
\rangle^2
\label{eq:BE2}
\end{equation}
where $\langle I_1K20|I_2K \rangle$ is a Clebsch-Gordan coefficient,
while $Q_0$ expresses the intrinsic electric quadrupole moment.  
The value $\beta \approx
0.4$ is extracted from the $Q_0$ value which is obtained by analyzing measured
strong E2 transitions within a given rotational band.  Thus, the extracted 
$\beta$-value
is consistent with that obtained from the comparison of the assigned 
$K^{\pi}$ values of the observed 
bands with the $\Omega^{\pi}$ values of one-particle orbits 
in the Nilsson diagram of Figure 7. 
The analysis of other available data, 
M1/E2 transitions, $\beta$ decays and  
one-nucleon transfer reactions, has been carried out in a similar manner
and provides further support to the interpretation based on the
Nilsson diagram  
\cite{BM75}. 

The analysis of available experimental data on medium-heavy deformed 
odd-A nuclei based on the Nilsson diagram, 
in which the pair correlation becomes important
and has to be taken into account in addition to the
spheroidal potential, has been even more quantitatively successful.  In those
heavier nuclei the notion of 
the deformed mean field is theoretically better justified; however, 
due to the higher density of one-particle levels the Nilsson diagram 
appropriate for those nuclei becomes more 
complicated.  
A beautiful example of observed data is
shown in Figures 8 and interpreted in terms of the Nilsson diagram given in
figure 9.  We refer the reader to many other such examples 
in the text book \cite{BM75}. 

\section{The yrast line}
The yrast line is defined by the quantum state with 
the lowest energy for a given angular momentum. 
In the yrast region the nucleus is cold in the sense that the entire
excitation energy of the nucleus is exhausted in generating the large total 
angular momentum.
Therefore, the structure in the yrast region is expected to be 
ordered with simple excitation
modes characteristic of the approach to zero temperature, 
and the study of quantal spectra in this region may be expected 
to give important
nuclear-structure information on how the nucleus responds to the large
centrifugal forces associated with rotation.

The path that the yrast line of actual nuclei will follow in deformation space
with increasing angular momentum will result from the interplay of the
macroscopic centrifugal distortion effect and quantal effects associated with
shell structure.  
A typical example of quantal effects is the observation of superdeformed bands 
($\delta \approx 0.6$) in nuclei with certain proton and neutron numbers along
the high-spin yrast line \cite{PT86}.  
With the singular exception of $^{8}$Be superdeformation has not been seen in
any nuclear ground state, but along the yrast line, the exceptionally large
moment of inertia associated with these large deformations provides the
explanation for the extensive regions where superdeformation is observed at
high angular momentum on the yrast line \cite{JK91}. 
The yrast configurations in some nuclei accommodate the angular momentum by
collective motion giving rise to regular sequences of levels (along a band
as in molecules),
while those in other nuclei correspond to rearrangements of the orbits of 
single or a few nucleons (involving a transition to a new band). 
The shape change along the yrast line and the related phenomena, 
which have been explored, 
may be conveniently found in various
review articles \cite{GHH86} and conference proceedings \cite{CONF}. 

The maximum value of the angular momentum that can be accommodated by a nuclear
system is limited by the
fission instability (in heavier nuclei) or by the ejection of particles 
carrying large orbital angular momentum (in lighter nuclei).\footnote{A very
different collective mode for generating states with large angular momentum    
is provided by the possibility of creating vortices in the quantum fluid. 
However, already a single vortex carries an angular momentum, $\hbar L$, of
order of the number of particles, $A$, and thus can only occur in finite
systems that are constrained by external force fields, as for example in the
high angular-momentum states of electrons in atoms.}  
  
When the coupling scheme along the yrast line is such that the angular momentum
is oriented approximately in the direction of the symmetry axis of oblate shape,
each one-particle orbit contributes a definite angular momentum in the direction
of the rotation axis.  Then, the transitions along the yrast line involve
successive rearrangements in the occupation of one-particle orbits.  In other
words, the collective moments of inertia are so small that each band contributes
only a single state to the yrast sequences.  Consequently, the energies on the
yrast line may exhibit considerable irregularity, and the transitions are at
most of one-particle strength and may suffer considerable 
hindrance which may
lead to the occurrence of yrast traps (yrast isomers). 

The manner of rotation together with the intrinsic deformation in a band 
may change as $I$ increases.  
As rotation sets in, the yrast shape of nuclei may become triaxial 
({\it i.e.\/} $R_x \neq R_y \neq R_z$) at a critical rotational frequency, and 
rotational bands may terminate at some angular momentum as 
in the $SU_{3}$ model, 
when the angular momentum $I$ is built up by a finite number of 
nucleons in a given open shell.
Approaching to the termination of the band is exhibited by a decrease of the
probabilities of E2 transitions within the band, since the gradual alignment 
of the nucleons leads towards a density distribution that is more and more 
symmetric 
about the axis of rotation \cite{BM75}.  
On the other hand, bands may not terminate if excited configurations
start to mix smoothly into rotational levels as $I$ increases.  
The latter has been shown to occur 
when the deformation of the band is larger than some
critical value \cite{TA79}.

\section{Miscellaneous}

We mention briefly seven topics 
concerning shape deformations which were not included 
in the previous sections: 
(i) the magic numbers (N, Z = 8, 20, 28, 50, 82, 126, ...) traditionally 
known in stable
nuclei may change in nuclei away from the stability line.  The change in
drip-line quantum-numbers comes from the presence of weakly-bound nucleons
(especially neutrons due to the absence of a Coulomb barrier), which leads 
to a change in shell-structure 
around the Fermi level. 
The neutron drip line has so far been experimentally reached 
in the oxygen (Z=8) isotope,
 for which $^{24}_{8}$O$_{16}$ is the heaviest element inside the drip line.  
Some neutron-drip-line nuclei
with a traditional magic number of neutrons such as N = 8 and 20 (for example, 
$^{12}_{4}$Be$_{8}$ \cite{HI00} and $^{30}_{10}$Ne$_{20}$ \cite{YY03}) 
are indeed 
found to have prolate ground-state deformation. 
(ii) It is now widely observed 
that a given nucleus accommodates various shapes
depending on excitation energies and spins. 
Even-even nuclei with closed-shell ground states exhibit excited states 
with strongly deformed
shape and rotational spectra. For example, $^{16}_{8}$O$_{8}$ (the rotational 
band beginning at 6.05 MeV) and $^{40}_{20}$Ca$_{20}$ (at 5.21 MeV) \cite{EI01}. 
(iii) The observation of wobbling mode in the neutron-deficient nucleus 
$^{163}_{71}$Lu$_{92}$ at moderate spins \cite{WOBL,DRJ02}
manifests the presence of triaxial quadrupole
deformed shape in nuclei. 
The characteristic feature of the wobbling mode has been pinned down by the
detailed study of the electromagnetic properties.
(iv) The observed almost complete dominance of prolate over oblate
deformations in the ground states of deformed even-even nuclei is not yet 
adequately understood.  
(v) The observation of remarkably low frequency negative-parity excitations
in some even-even actinide nuclei (for example, 
the $I^{\pi} = 1^-$ state observed
at 0.216 MeV in $^{224}_{88}Ra_{136}$ \cite{SAP54}) 
indicates an incipient octupole
instability in this nucleus.   
(vi) It is remarked that presently available self-consistent mean-field
approximation to the nuclear many-body Hamiltonian with effective interactions
gives a good description of the observed regions of 
quadrupole-deformed nuclei \cite{DV73}, though observed prolate dominance is
unresolved.  For a discussion of the quadrupole correlation effect described by
the generator coordinate method, see \cite{BEN06a}.  On the other hand, 
shell-model calculations of conventional type are not quantitatively successful 
in studying the
properties of deformed nuclei except for light nuclei, 
basically because the
configuration space must be truncated 
in order to be able to carry out the calculation.  
A recent development of the shell-model technique in the
study of deformed nuclei can be found in \cite{EC05}.  
(vii) In a quantal system such as nuclei, the definition
of deformation requires the condition that the zero-point shape fluctuations are
small compared with the equilibrium deformation values.  The distribution of the
ratio $R = E(4^{+}_{1}) / E(2^{+}_{1})$ observed in even-even nuclei has a sharp
peak around $R = 3.3$, which indicates an axially-symmetric 
rotor with a very weak coupling to
intrinsic motion.  Available systematic microscopic
calculations suggest the relatively rigid shape of these deformed
nuclei \cite{JPD10}.  
 
As a final remark, we may mention; (a) the construction of facilities providing 
radioactive ion beams has in recent years made (and will make) it possible to
reach out to the region of the nuclear chart further away from stability;  
(b) the remarkable development of the advanced $4 \pi$ $\gamma$-ray detector
systems has led to the observation of weakly-populated states of exotic shapes
and/or extreme angular momentum.

\newpage

{\bf\large Figure captions}\\
\begin{description}
\label{fig:n-p-def}
\item[{\rm Figure 1 :}]
Regions of deformed nuclei.
The nuclei, for which both $N$ and $Z$ are even numbers 
(called even-even nuclei), have ground-state spin-parity 0$^{+}$, 
without exception. 
The overwhelming majority of these has 2${^+}$ first excited state. 
Writing the excitation energies of the lowest-lying 2$^+$ and 4$^+$ states 
as $E(2_1^+)$ and $E(4_1^+)$, the 
filled circle represents even-even nuclei, in which 
E(4$^+_1$)/E(2$^+_1$) $>$ 2.7. 
The data are taken from http://www.nndc.bnl.gov/ensdf/.
The line of $\beta$ stability is indicated by the thin long-dashed curve. 
The thin straight lines parallel to the x and y axes 
show the magic numbers of protons and neutrons, which
are known in nuclei along the $\beta$ stability line. 
Except for very light nuclei (Z $\leq$ 8) the neutron drip line, 
at which nuclei become unstable for neutron emission, is not known
experimentally.  The boarder of deformed nuclei shown for 
the neutron-rich region of medium-heavy nuclei is often equal to the boarder
of neutron-rich nuclei, for which the energy of the 4$^+_1$ state is presently 
known.
The criterion of deformed even-even nuclei can be made 
in one of the following four ways;
(i) the excitation spectra exhibit an approximate $I(I+1)$ energy dependence
indicating rotational structure.  In the figure the (rather arbitrary) 
criterion E(4$^+_1$)/E(2$^+_1$) $>$ 2.7 is chosen. 
(ii) B(E2) values are
much larger than single particle or collective vibrational estimates.  
(iii) The excitation energy of the first excited
2$^+$ state is especially low, say by a factor of at least 5, 
compared with twice
the odd-even mass difference, $2 \Delta$.  We note that 
the ratio, $E(2^+_1) / 2 \Delta$, has 
the mass-number dependence of about $A^{1/6}$.  
(iv) The ratio, $Q(2^+_1)^2/B(E2;2^+_1 \rightarrow 0^+_{gr})$, where $Q(2_1^+)$ 
expresses the quadrupole moment of the lowest 2$^+$ state while $B(E2)$ denotes
the reduced probability of electric-quadrupole transitions, is 
approximately equal to 4.10, 
which is the value for the ideal collective rotation.
\end{description}

\begin{description}
\item[{\rm Figure 2 :}]
Intrinsic and laboratory fixed coordinate systems.  The Euler angles ($\phi,
\theta, \psi$) are denoted, collectively, by $\bf\Omega$. 
The curve expresses symbolically the intrinsic body, for which the deformation
completely breaks the rotational symmetry.
\end{description}

\begin{description}
\item[{\rm Figure 3 :}]
Rotational coupling scheme describing rotational motion of a spheroidal
shape.
The symmetry axis is labeled by $S$, the total angular momentum 
by ${\bf I}$ and the 
collective rotational angular-momentum by ${\bf R}$, while $K$ is the projection
of ${\bf I}$ on $S$.  Since $\bf R$ must be perpendicular to $S$, $K = \Omega$, 
where $\Omega$ is associated with the intrinsic state $\phi_{K}$. 

\end{description}

\begin{description}
\item[{\rm Figure 4 :}]
(a) Observed rotational bands in $^{166}_{68}$Er$_{98}$. 
Excitation energies (in MeV) of levels are written on the r.h.s., while the
spin-parity $I \pi$ on the l.h.s.  The levels belonging to a band are connected
by strongly enhanced E2 transitions.  
The bands are labeled by the component 
$K^{\pi}$ of the total angular momentum with respect to the symmetry axis
;   
$K^{\pi} = 0^+$ and 2$^+$ for the ground and the first excited bands, 
respectively.  
For $K$=0 $R$-symmetry implies 
a restriction on the I-values 
fulfilling the condition,  
($-$1)$^I$ = $r$.  Consequently, only 
even- or odd-$I$ values appear in a given rotational band. 
The ground state band of deformed even-even nuclei
consists of even integer values of $I$ with positive parity, 
since the configuration, 
in which a doubly degenerate time-reversed pair of one-particle levels 
in a deformed potential are pairwise filled, has $r$=+1.  
(b) Three lowest-lying rotational bands are observed in $^{238}_{92}$U$_{146}$. 
Due to $R$-symmetry of the deformation the rotational band with $K^{\pi}$=$0^-$ 
and $r=-1$ has members with $I^{\pi} = 1^-, 3^-, 5^-$, ..., while the one with 
$K^{\pi}$=$0^{-}$ and $r=+1$ has those with $I^{\pi} = 0^-, 2^-, 4^-,$ ... 
\end{description}

\begin{description}
\item[{\rm Figure 5 :}]
Single-particle spectrum for an axially-symmetric harmonic-oscillator 
potential. 
The energy eigenvalue of the one-particle orbit ($n_x, n_y,
n_z$) is written as 
\begin{equation}
\varepsilon (N,n_z) = \hbar \bar{\omega} \left( N + \frac{3}{2} - \frac{1}{3} 
\delta_{osc} (3n_z-N) \right)  
\label{eq:dhospe}
\end{equation} 
where the principal quantum number $N = n_{\perp} (\equiv n_x + n_y) + n_z$ 
and $\bar{\omega} = (2 \omega_{\perp} + \omega_z) /3$, 
while the deformation parameter $\delta_{osc}$, which is
approximately equal to $\beta \approx \delta$,  is defined by  
\begin{eqnarray*}
\delta_{osc} = 3 \, \frac{\omega_{\perp}-\omega_z}{2\omega_{\perp} + \omega_z} 
\approx \frac{R_z - R_{\perp}}{R_{av}}  
\end{eqnarray*}
where $R_{av}$ denotes the mean radius. 
At $\delta_{osc} = 0$ (spherical shape) the spectrum is 
regularly 
bunched with the equal energy spacings $\hbar \omega_0$.  
Each major shell with the
energy $(N+\frac{3}{2}) \hbar \omega_0$ has the degeneracy $(N+1)(N+2)$
including the nucleon spin degree of freedom.  
The particle numbers, 2, 8, 20, 40, 70, 112, ..., would be the magic
numbers for this potential and associated 
with an especially stable spherical shape. 
Single-particle levels belonging to a given major shell have the same parity
$\pi = (-1)^N$ with the maximum orbital angular momentum $\ell_{max} = N$.  
For $\delta_{osc} \neq 0$ the levels split into $(N+1)$ levels with eigenvalues
$\varepsilon (N, n_z)$ in (\ref{eq:dhospe}), and each level has 
a degeneracy of 2($n_{\perp} + 1$), 
where a factor 2 comes from the nucleon spin $\frac{1}{2}$ 
while ($n_{\perp} + 1$) from
possible values of $n_x$($= 0, 1, 2, ..., n_{\perp}$).  
For prolate (oblate) deformation,
$\omega_z < \omega_{\perp}$ or $\delta_{osc} > 0$ 
($\omega_z > \omega_{\perp}$ or $\delta_{osc} < 0$), the levels with larger
(smaller) $n_z $ become energetically lower. 
Eigenvalues expressed in units of the mean frequency $\bar{\omega}$ have 
a linear dependence on $\delta_{osc}$, and the slope of the lowest level with a
given $N$ for prolate shape ($n_z = N$) is twice that for oblate shape 
($n_z$ = 0) with an opposite sign.  
The arrows mark the deformations corresponding to the indicated rational ratios
of frequencies $\omega_{\perp} : \omega_z$.  
The figure gives the total particle number
corresponding to completed shells in the potentials 
with $\omega_{\perp} : \omega_z$ 
= $1 : 1$ (spherical), $2 : 1$ (prolate) and $1 :2$ (oblate).  
The figure is taken from Ref. \cite{BM75}.
\end{description}

\begin{description}
\item[{\rm Figure 6 :}]
Observed low-lying level scheme of $^{25}$Mg. 
Excitation energies of levels are expressed in units of MeV. 
The levels are mostly obtained from the reaction of $^{24}$Mg
(d, p $\gamma$)$^{25}$Mg and, consequently, belong to 
the $^{24}$Mg plus one-neutron configurations. In other words, 
only the one-particle levels lying higher 
than the [202 5/2] level for prolate shape in th Nilsson diagram of Figure 7 
are surely observed.
The parameters in the expression (\ref{eq:roten}), 
$A$ for $K \neq 1/2$ bands ($A$ and $a$ for $K$=1/2 bands), 
which are estimated using the energies of 
the lowest two (three) rotational members, are 
indicated below respective bands. 
It is seen that already around band heads the rotational spectra of
$K$=1/2 bands often deviate considerably from the $I(I+1)$ dependence. 
The decoupling parameters, which are estimated using respective $\Omega = 1/2$
particle wave functions in the spheroidal potential with $\beta \approx 0.4$, 
exhibit the characteristic behavior depending on different bands and are found 
to provide a further support for the assignment of the intrinsic configurations
given in the figure \cite{BM75}.  

\end{description}

\begin{description}
\item[{\rm Figure 7 :}]
Neutron one-particle levels in a spheroidal Woods-Saxon plus spin-orbit 
potential, for which 
the radius $R$ and the depth $V_{WS}$ are 
adjusted approximately to those of $^{25}_{12}$Mg$_{13}$. 
The parameters of the potential at $\beta$=0 are taken from p.239 of Ref.
\cite{BM69}. 
Positive-parity one-particle levels are denoted by solid curves.
One-particle levels with $\varepsilon_{\Omega} > 0$ are obtained as one-particle
resonant levels in the potential.  The associated resonant widths are not shown,
for simplicity.  
Calculated one-particle resonant energies in MeV at $\beta$=0 are : 
$\varepsilon_{res}$(p$_{3/2}$)=0.32, $\varepsilon_{res}$(f$_{7/2}$)=0.33, 
and $\varepsilon_{res}$(f$_{5/2}$)=10.37, while the p$_{1/2}$ level 
expected just above the $p_{3/2}$ and $f_{7/2}$ levels is not
obtained as a resonant level for the present potential. 
Bound one-particle levels are 
labeled by the asymptotic quantum numbers [$N n_z \Lambda \Omega$], 
where $\Omega$ represents the component of the total angular
momentum along the symmetry axis and is a constant of the motion for all values
of $\beta$.  The additional quantum numbers refer to the structure of the orbits
in the limit of large deformations : the total number of nodal surfaces ($N$),
the number of nodal surfaces along the symmetry axis ($n_z$), and the component
of orbital angular momentum along the symmetry axis ($\Lambda$).  Each level is
doubly degenerate ($\pm \Omega$) associated with time reversed invariance of the
Hamiltonian.
The neutron numbers 16 and 20, which are obtained by filling in all lower-lying
levels, are indicated with circles.
A pairwise filling of the doubly degenerate ($\pm \Omega$) orbits contributes
no net angular momentum along the symmetry axis.  
For example, in $^{25}$Mg$_{13}$ with the neutron number 13 
it is seen that the lowest two intrinsic configurations for 
$\beta \approx 0.4$ are expected to involve the 13th neutron occupying the
orbits [202 5/2] and [211 1/2], in agreement with the observed spectroscopic
properties shown in Figure 6.  
The observed second and
third excited bands starting at 2.563  and 3.413 MeV (Figure 6) can be
interpreted as having the [200 1/2] and [330 1/2] intrinsic configurations,
respectively, for the odd neutron. 
\end{description}

\begin{description}
\item[{\rm Figure 8 :}]
Observed low-lying rotational bands of $^{175}_{70}$Yb$_{105}$. 
Excitation energies of levels are expressed in units of MeV. 
The levels which are observed by (d,p) reactions and thus classified 
as particle states are drawn to the right of the ground-state band [514 7/2], 
while hole states observed by (d,t) reactions are drawn to the left. 
Only a few lowest-lying members are plotted for each rotational band, for
simplicity.  The classification of the intrinsic states is supported by the
observed intensities in one-neutron transfer reactions, observed rotational
energies and transition probabilities.  See Ref. \cite{BM75} for the detailed
analysis. 
The data are taken from Nuclear Data sheets, {\bf 102} (2004) 719 and 
http://www.nndc.bnl.gov/chart/.
\end{description}

\begin{description}
\item[{\rm Figure 9 :}]
Neutron one-particle levels in a spheroidal Woods-Saxon 
plus spin-orbit potential,
for which the radius $R$ and the depth $V_{WS}$ are adjusted approximately to
those of $^{175}_{70}$Yb$_{105}$.  The parameters of the potential at $\beta$=0
are taken from p.239 of Ref. \cite{BM69}, except the strength of the spin-orbit
potential that is about 20 percent stronger here.  The neutron numbers, 
82, 104 and 126, which are obtained by filling in all lower-lying levels, are
indicated with circles.  It is seen that in the deformation region 
of $0.15 < \beta < 0.36$ the 105th neutron occupies 
the [514 7/2] level, in agreement with the observed 
ground-state configuration of $^{175}$Yb.  Moreover, the observed sequence of
intrinsic states shown in Figure 8 is seen to be consistent with the
single-particle spectrum in the present figure at $\beta \approx 0.28$: in
particular, 
above the [514 7/2] level we find the [624 9/2], [510 1/2], [503 7/2], 
[512 3/2] and [651 1/2] levels, while below the [514 7/2] level we have 
the [512 5/2], [633 7/2] and [521 1/2] levels. 
In the figure only the one-particle levels, which lie in the neighborhood of 
the [514 7/2] level at $\beta$=0.28, are marked by respective asymptotic 
quantum numbers [$N n_z \Lambda \Omega$]. 

\end{description}

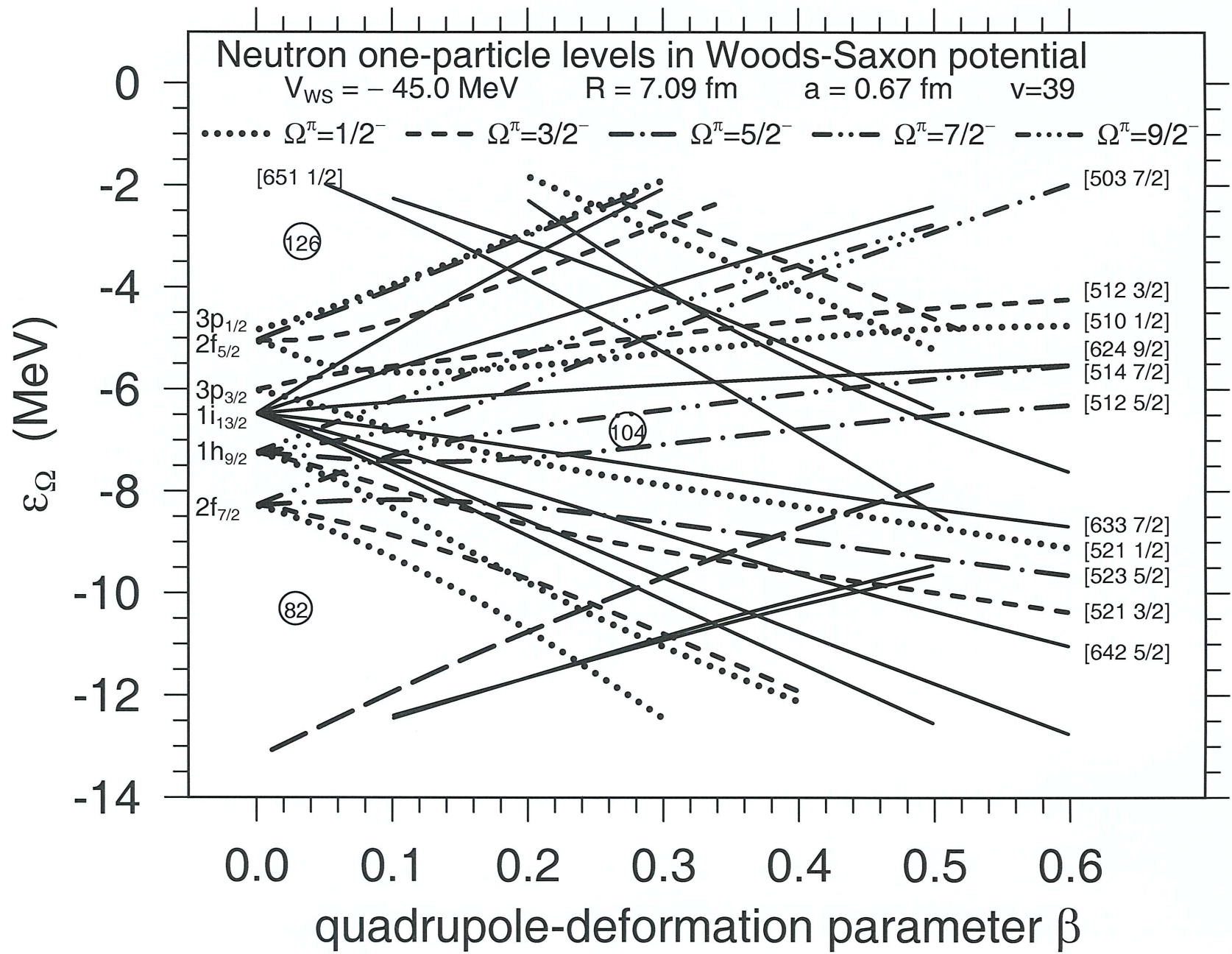
\end{document}